\documentclass[prl,twocolumn,preprintnumbers]{revtex4-1}
\pdfoutput=1

\usepackage{graphicx}  
\usepackage{amssymb,amsmath}   
 \usepackage{url}
 \usepackage{hyperref}
 \hypersetup{colorlinks=true,linkcolor=red,anchorcolor=green,citecolor=blue,filecolor=black,menucolor=black,urlcolor=black}

\begin{document}

 \preprint{  	DAMTP-2017-38 }

 \title{Windings of twisted strings}
  
\author{Eduardo Casali$^\dagger$, Piotr Tourkine$^{\ddagger}$
\vspace{.1cm}
\\ \small{$^\dagger$ The Mathematical Institute, University of Oxford, Woodstock Road, Oxford OX2 6GG, UK
\\ $^\ddagger $ CERN, Theory Group, Geneva, Switzerland
}}

\email{eduardo.casali@maths.ox.ac.uk,piotr.tourkine@cern.ch}

\begin{abstract}
  Twistor string models have been known for more than a decade now but have
  come back under the spotlight recently with the advent of the scattering equation formalism which has greatly generalized the scope of these models.
  A striking ubiquitous feature of these models has always been that,
  contrary to usual string theory, they do not admit vibrational
  modes and thus describe only conventional field theory.
  In this paper we report on the surprising discovery of a whole new
  sector of one of these theories which we call ``twisted strings'', when spacetime has compact directions. We find that the
  spectrum is enhanced from a finite number of states to an infinite
  number of interacting higher spin massive states.
We describe both bosonic and worldsheet supersymmetric models, their spectra and scattering amplitudes.
These models have distinctive features of both string and field theory, for example
they are invariant under stringy T-duality but have the high energy
behaviour typical of field theory.
Therefore they describe new kind of field theories in target space,
sitting on their own halfway between string and field theory.

\end{abstract}

\maketitle

\section{Introduction}
\label{sec:introduction}

String theories based on Penrose's twistor theory have lead to a
revolution in our understanding of the S-matrix of quantum field
theory~\cite{Witten:2003nn}. The most remarkable feature of these models is that, contrary to string
theory, they describe only field theory (spin $\leq$ 2 fields), that is, higher spin massive excitations are absent from their spectrum.
These models combine the mathematical elegance
of string theory with twistor methods to describe the low energy field theories of nature.

In the field of scattering amplitudes, the recent introduction of the ``scattering equations
formalism''~\cite{Cachazo:2013iea,Cachazo:2013hca,Cachazo:2013gna}
lead to the discovery of the ambitwistor string, which in turn lead
to advances which were out of reach of the previous twistor methods.
Most notably this includes
loop-level~\cite{Adamo:2013tsa,Adamo:2015hoa,Ohmori:2015sha,Geyer:2015bja,Geyer:2015jch}
amplitudes and curved space~\cite{Adamo:2014wea,Adamo:2017sze}. But an
important open question remained, the connection between these new
models to traditional string theory.
The best framework to understand these questions relies on a recently
re-discovered quantization ambiguity that leads to closed theories which we
call ``twisted strings'' below. The idea is that the ambitwistor string arises in the \textit{tensionless
  limit} of the twisted
string~\cite{Siegel:2015axg,Casali:2016atr,Lee:2017utr,Azevedo:2017yjy},\footnote{This is counter-intuitive because field theories are also
  found in the infinite tension limit of conventional string
  theory. Note also that we choose the terminology ``twisted'' instead
  of ``chiral'' or ``left-handed'' as first used
  in~\cite{Huang:2016bdd,Siegel:2015axg} because our string is not chiral.}.

The main result of this paper is the following: While in flat
space the twisted type II string describes gravity only, in spacetimes
with different topologies, where the string can wind around compact
dimensions, an infinite tower of massive states arises. This
twisted string theory remains distinct from usual string theory, as we explain
below, and describes new type of theories halfway
between string and field theory.

We study the bosonic and worldsheet supersymmetric models. The bosonic
twisted string has tachyonic excitations (whose masses squared are
bounded by $-4/\alpha'$ where $\alpha'$ is the Regge slope) and ghosts
(negative normed states) in the physical spectrum. The
Ramond-Neveu-Schwarz (RNS) formulation with $\mathcal{N}=(1,1)$
worldsheet supersymmetry has no tachyons but also seems to have
non-decoupling ghosts.

Remarkably, we also find that the compactified theory has T-duality:
the spectrum is invariant when the radius $R$ of the compact dimension
is exchanged with $\alpha'/R$. At the self-dual radius
$R=\sqrt{\alpha'}$, both in the RNS and bosonic models, infinitely
many higher spin states become massless, a rather striking fact that
will be explored elsewhere.

We compute the scattering amplitude of these twisted theories and find
that, in contrast to string theory, these theories do \textit{not} enjoy
exponential suppression at high energies~\cite{Amati:1987wq,Gross:1987ar}.

The combination of these three elements: stringy T-duality,
field-theory power law suppression at high energy and the fact
that these theories only describe conventional field theory in
flat space give these theories a genuinely novel
status somewhere between field and string theory.

We also comment on the connection to the ambitwistor
string~\cite{Mason:2013sva} and scattering equations in the
tensionless limit~\cite{Siegel:2015axg,Casali:2016atr}. We find that
the winding modes decouple from the scattering equations, but modify
the integrand in a way that we describe. This gives new scattering
equation-like formulas for winding states which include
higher-spin states.

\section{Twisted strings}
\label{sec:twisted-strings}

\paragraph{Review:} We call twisted strings the worldsheet models
described by Siegel in the context in~\cite{Siegel:2015axg}, \footnote{See also the older work
  \cite{Hwang:1998gs}.}. The $\alpha'\to\infty$ limit
of these models produce the ambitwistor
string~\cite{Siegel:2015axg,Casali:2016atr,Lee:2017utr}. We 
present here a brief description of the model, for more details the reader is referred to \cite{Lee:2017utr}.
The twisted bosonic string action is the Polyakov action
\begin{equation}
 S=\frac{1}{2\pi\alpha'}\int d^2z \partial X^\mu \bar\partial X_\mu
\end{equation}
here written in flat space and in conformal gauge. We expand the field
$X(z,\bar{z}) = X_L(z) + X_R(\bar{z})$ in the usual way
\begin{equation}
 \begin{aligned}  
  X_L^\mu(z)&=x_L^\mu -i \frac{\alpha'}{2}p_L^\mu \ln(z) + i
          \left(\frac{\alpha'}{2}\right)^{1/2}\sum_{m=-\infty}^{\infty} \frac{1}{m}\frac{\alpha_m^\mu}{z^m}\\
  X_R^\mu(z)&=x_R^\mu -i \frac{\alpha'}{2}p_R^\mu \ln(\bar z) + i
          \left(\frac{\alpha'}{2}\right)^{1/2}\sum_{m=-\infty}^{\infty}
          \frac{1}{m}\frac{\tilde \alpha_m^\mu}{\bar z^m}
 \end{aligned}
\label{eq:field_exp}
\end{equation}
with canonical commutation relations
\begin{align}
[\alpha_n^\mu,\alpha_m^\nu] =[\tilde\alpha_n^\mu,\tilde\alpha_m^\nu]=
n \delta_{n+m} \eta^{\mu\nu}\label{eq:com-rel-bos}\\
  \label{eq:zm-comrel}
  [x_{L/R}^\mu,p_{L/R}^\nu]=i \eta^{\mu\nu}\,,\quad   [x_{L/R}^\mu,p_{R/L}^\nu]=0
\end{align}
where $\eta^{\mu\nu}=(-,+,\dots,+)$.

The difference to conventional strings comes from the choice of
vacuum~\cite{Bagchi:2015nca,Siegel:2015axg,Casali:2016atr,Hwang:1998gs,Lee:2017utr}
\begin{equation}
  \label{eq:vacuum}
  \alpha_n |0\rangle = \tilde \alpha_{-n}|0\rangle = 0,\quad \forall n>0,
\end{equation}
which we call \textit{twisted vacuum}. The negative modes of
$\tilde\alpha$ annihilate the vacuum in contrast to the conventional string
vacuum where the positive modes annihilate the
vacuum. This defines the following operator ordering
\begin{equation}
  \label{eq:normal-ordering}
  :\alpha_n^\mu \alpha_{-m}^\nu: = \alpha_{-m}^\nu \alpha_{n}^\mu,
  \quad
  :\tilde \alpha_{-m}^\mu \tilde\alpha_{n}^\nu:  = \tilde
  \alpha_n^\nu\tilde \alpha_{-m}^\mu\,,\quad \forall n,m>0\,.
\end{equation}

The spectrum of these theories can be computed from the Virasoro
constraints
\begin{equation}
(\partial X)^2=0, \quad (\bar \partial X)^2=0\,.
  \label{eq:constraints}
\end{equation}
The bosonic theory is found to live in 26 dimensions. In the
twisted vacuum the zero modes $L_0$ and $\bar L_0$ of \eqref{eq:constraints} acquire a normal ordering constant~\footnote{See \cite{Lee:2017utr} for more details on these expressions.},
\begin{align}
  \label{eq:L0}
  L_0 =\tfrac{{\alpha'}}4 p_L^2 + N - 1\,,\quad \bar L_0 = \tfrac{{\alpha'}}4 p_R^2 - \bar N +  1
\end{align}
with $N = \sum_{n=1}^\infty:\alpha_{-n}\cdot\alpha_n:$ and
$\bar{N} = -\sum_{n=1}^\infty:\tilde\alpha_{-n}\cdot\tilde\alpha_n:$ -- the minus
signs comes from the choice of vacuum.
The zero-modes of the Virasoro conditions then read
\begin{equation}\label{eq:zero_modes}
(N-\bar{N})-\alpha' m^2=0;\;\;\; N+\bar{N}-2=0
\end{equation}
The normal ordering constant appears in the level matching truncating the spectrum to a finite number of
states. These are the massless sector of the
bosonic string plus massive spin two states, with $m^2 = \pm 4/\alpha'$.

Adding a pair of fermions gives the RNS model which also has a twisted
vacuum (see eq.~\eqref{eq:fermion-twisted}) but in
this case there is no tachyon. We proved in \cite{Casali:2016atr}
following on~\cite{Siegel:2015axg} that both the bosonic and the type
II models have as the tensionless limit ambitwistor strings.

\medskip

\paragraph{Bosonic model on a circle:}
\label{sec:bosonic-model}
Taking one of the target-space dimensions to be a circle of radius $R$
allows winding modes $X^{25}(\sigma+2\pi)=X^{25}(\sigma)+2w\pi R$ with
$w\in\mathbb{Z}$. In the presence of winding and Kaluza-Klein modes
($n\in\mathbb{Z}$) in the 25th dimension, the momentum zero modes are
\begin{equation}
  \label{eq:mom-def}
  p_{L/R}^\mu = \left(k^m,\frac{n}{R}\pm \frac{R w}{\alpha'}\right)
\end{equation}
where $m=0,\dots,24$.
The constraints \eqref{eq:zero_modes} become
\begin{equation}
  \begin{aligned}
  &m^2 = \frac {n^2}{R^2} + \frac{ R^2 w^2}{\alpha'^2} + \frac{2}{\alpha'}(N-\bar N)\\
  &N+\bar N +n w =2.
\end{aligned}
\label{eq:Vir-S1}
\end{equation}
The compactified model contains many new tachyonic states. To see this
take $k=-nw>0$. The lowest possible mass squared states have
$(N,\bar{N})=(0,k+2)$. The mass-shell constraint then gives
\begin{equation}
 \begin{aligned}
   m^2 =& \frac {n^2}{R^2} + \frac{ R^2 w^2}{\alpha'^2} + \frac{2}{\alpha'}(-k-2) = \left(\frac{n}{R}+\frac{wR}{\alpha'}\right)^2 -\frac{4}{\alpha'}
 \end{aligned}
\label{eq:bosonic_tachyons}
\end{equation}
Taking $n>0$ and $w<0$, $m^2<0$ if
\begin{equation}
  n > \frac{|w|R^2}{{\alpha'}}-2\;\;\text{and}\;\; n < \frac{|w| R^2}{{\alpha'}}+2
\end{equation}
There are always integer
solutions for $n$, therefore there are infinitely many tachyonic
states in the bosonic twisted string, with masses squared bounded from
below by $-4/\alpha'$. In addition, many of them have negative
norm \cite{Leite:2016fno,Lee:2017utr}, we come back to this point for the RNS model below.
\medskip
\paragraph{RNS model on a circle:}
\label{sec:rns-model}
The RNS model contains two Majorana-Weyl fermions
$\psi(z),\bar \psi(\bar z)$ and has the standard RNS worldsheet
supersymmetric action. Following
conventions of~\cite{Polchinski:1998rr} the fermions have mode
expansion
\begin{align}
  \label{eq:fermions}
  \psi(z)=\sum_{r\in\mathbb{Z}+\nu} \frac{\psi_r}{z^{r}},\qquad
  \bar \psi(\bar z)=\sum_{r\in\mathbb{Z}+\tilde \nu} \frac{\bar\psi_r}{z^{r}}
\end{align}
with $\nu=1/2$ is the NS sector while $\nu=0$ is the
Ramond sector, and we keep
\begin{equation}
  \label{eq:comm-rel-psi}
  \{\psi_r^\mu,\psi_s^\nu\} =  \{\tilde\psi_r^\mu, \tilde\psi_s^\nu\} =\eta^{\mu\nu}\delta_{r+s}
\end{equation}
while we define the twisted vacuum
\begin{equation}
  \label{eq:fermion-twisted}
  \psi_{r}^\mu|0\rangle =0\,,\quad \tilde
  \psi_{-r}^\mu|0\rangle=0\quad \forall r>0
\end{equation}
depending on the NS or R vacuum.
In a given sector for the closed string, NS-NS, NS-R, R-NS or R-R, the constraints are
\begin{equation}
  \begin{aligned}
  &m^2 = \frac {n^2}{R^2} + \frac{ R^2 w^2}{\alpha'^2} +
  \frac{2}{\alpha'}(N-\bar N-a^\nu+a^{\bar \nu})\\
  &N^{tot}+\bar N^{tot} +n w =a^\nu+a^{\bar \nu}
\end{aligned}
\label{eq:Vir-S2}
\end{equation}
with $a^{NS} = 1/2,a^R = 0$. The operators $N^{tot}$ and $\bar
N^{tot}$ count the total number of oscillators, $\psi$'s and $\alpha$'s.

These models have a Gliozzi-Scherck-Olive (GSO) projection which
eliminates the tachyonic state. The argument is: In the NS-NS sector,
the GSO-even states are built out of at least one $\psi^\mu_r$ and
$\tilde{\psi}^\nu_r$ modes, for example
$\psi_{-1/2}^\mu\psi^\nu_{1/2}|0\rangle$. Therefore the number
operators on physical states always obey $N^{tot}\ge\frac{1}{2}$ and
$\bar{N}^{tot}\ge\frac{1}{2}$.

It is easy to see that there are no tachyonic modes in the NS-NS
sector; the only non-trivial case has $nw=-k$ with $k>0$. The twisted
level matching constraint implies that the lowest mass in
\eqref{eq:Vir-S2} is a state with $\bar{N}^{tot}=k+\frac{1}{2}$ and
$N^{tot}=0$;
\begin{equation} \begin{aligned}
 m^2 = \frac {n^2}{R^2} + \frac{ R^2 w^2}{\alpha'^2} +
 \frac{2}{\alpha'}\left(\frac{1}{2}-k-\frac{1}{2}\right) 
 =\left(\frac{n}{R}+\frac{wR}{\alpha'}\right)^2\,,
\end{aligned}
\label{eq:RNS-lowes-mass}
\end{equation}
which is always positive.
In the R-R sector both $N^{tot}$ and $\bar{N}^{tot}$ can be zero but
there is no normal ordering constant. The lowest possible value for
the mass is given by the state with $(N,\bar{N})=(0,k)$ for which
\begin{equation}
\begin{aligned}
 m^2 = \frac {n^2}{R^2} + \frac{ R^2 w^2}{\alpha'^2} + \frac{2}{\alpha'}\left(-k\right)
 =\left(\frac{n}{R}+\frac{wR}{\alpha'}\right)^2>0.
\end{aligned}
\label{eq:RNS-lowes-mass-R}
\end{equation}
The mixed NS-R and R-NS sectors are also tachyons-free, the argument
follows as above. In conclusion, the compactified twisted model has no
tachyons.

\medskip

\paragraph{Negative-norm states:}
%
So far we have described the masses of the states in the spectra of
these new theories. It turns out that the models possess ghosts in the
physical spectrum, which naively spoils the unitarity of the
theories. In the flat space case, it was known that the bosonic model
had ghosts, while the supersymmetric model was ghost-free. Here this
is not the case anymore and we argue now that both models do contain
ghosts in the physical spectrum. In light-cone gauge, the GSO
admissible supergravity states are of the form
$\psi_{-1/2}^I \bar \psi_{1/2}^J |0,k\rangle$ with $I,J=1,...,d-1$
have positive norm if $\langle0,k|0,k\rangle>0$. A winding state with
KK momentum and winding is created by acting with
$\exp(i k_L \alpha_0)$ and $\exp(i k_R \tilde \alpha_0)$ on
$\psi_{-1/2}^I \bar \psi_{1/2}^J |0,k\rangle$, adding compensating
$\tilde{\alpha_i}$'s and setting the number of $\alpha$'s to zero such
that $\bar{N}^{tot},n,w$ solve the constraints. These states have
negative norms whenever there is an odd number of
$\tilde{\alpha}$~\footnote{Once the higher Virasoro constraints are
  applied $\bar L_{-m}|\mathrm{phys}\rangle=0$ with $m>0$, there still
  remain infinitely many physical negative-norm states, as can be
  easily checked.}. An interesting prospect would be to find a target
space with extra symmetries that would remove these states.

\paragraph{T-duality:}
\label{sec:t-duality}
Another fact to add to the list of curious properties of these new
theories is that they are invariant under T-duality
\begin{equation}
  \label{eq:T-duality}
  (n,w,R)\leftrightarrow (w,n,\frac{\sqrt{\alpha'}}{R}).
\end{equation}
This is surprising since in normal string theory, T-duality reflects
the existence of a minimum spacetime length and is connected to the UV
completeness of string theory. The amplitude analysis below shows
that, contrary string theory amplitudes that are exponential soft at
high energies, twisted strings have a power-law fall-off and therefore
behave like field theories, which do generically suffer from UV
divergences.

Since the action is the same for twisted strings and conventional strings we expect that the Buscher rules~\cite{Buscher:1987qj} for T-duality in non-trivial backgrounds are the same in twisted strings as in string theory.

At the self-dual radius $R=\sqrt{\alpha}'$ a surprising effect
arises: 
both the twisted bosonic and RNS strings do have infintely many extra
massless states at the self-dual radius.
The Virasoro conditions obeyed by physical states can be written as follows
\begin{equation}
\begin{aligned}
 m^2  = (k_L)^2 +\frac{4}{\alpha'}\left(N^{tot}-\frac{1}{2}\right)
 =
 (k_R)^2 +\frac{4}{\alpha'}\left(\frac{1}{2}-\bar{N}^{tot}\right)
 \end{aligned}\label{compact-mom}
\end{equation}
Since $N^{tot},\bar{N}^{tot}\ge\frac{1}{2}$ for most values of
$N^{tot}$ the first condition has no solution when $m^2=0$. However
there's an interesting class of solutions with
$N^{tot}=\frac{1}{2}$. The first equation is solved by $w=-n$ and the
second by $w=\pm \sqrt{q}$ with $\bar{N}^{tot}=\frac{1}{2}+q$. For
$\sqrt{q}$ integer, these are consistent solutions, that describe
massless spacetime vectors with non-zero KK and winding charge. This
suggests an exotic gauge symmetry
enhancement~\footnote{Note further that for rational values of
  $R/\sqrt{\alpha'}=p/q\in\mathbb{Q}$ infinitely many massless states
  are generated. This is more evidence of the unconventionality of the
  target space theory.}.
It would be nice to understand this strange sector of the theory. One
interesting application would be to take the tensionless limit of an
amplitude between these gauge bosons and obtain corresponding CHY
formulae. Since the heterotic ambitwistor string is
inconsistent~\cite{Mason:2013sva}, this may provide an alternative
approach to gauge interactions.

\section{Properties of twisted strings}
\label{sec:prop-twist-strings}

In this section we study various properties of the twisted strings
with windings; scattering amplitudes, high-energy behaviour and partition
functions. We also comment on the form of ambitwistor or CHY
integrands with windings.

\medskip

\paragraph{Scattering amplitudes and high-energy behaviour:}
\label{sec:ampl-high-energy}
It was shown in
\cite{Siegel:2015axg,Huang:2016bdd,Leite:2016fno,Lee:2017utr} that the
oscillator flip in scattering amplitudes is implemented by using the
twisted worldsheet correlators $\langle X X\rangle$ (see also
\cite{Tseytlin:1990va})
\begin{equation}
  \langle X^\mu(z,\bar z) X^\nu(w,\bar w)\rangle =- \frac{\alpha'}{2} \ln
  \left(\frac{z-w}{\bar z-\bar w}\right)
\label{eq:twisted-XX}
\end{equation}
or, equivalently,
\begin{align}
    \langle X_L^\mu(z) X_L^\nu(w)\rangle =- \eta^{\mu\nu}\frac{\alpha'}{2} \ln
  \left({z-w}\right)\label{eq:XlXl}\\
      \langle X_R^\mu(\bar z) X_R^\nu(\bar w)\rangle = \eta^{\mu\nu}\frac{\alpha'}{2} \ln
  \left({\bar z-\bar w}\right)\label{eq:XrXr}
\end{align}

The part of the string integrands that are affected by the sign
flip is the Koba-Nielsen factor, the correlator between the
plane-wave part of the vertex operators $\exp(i k\cdot X(z,\bar z))\to
\exp(i(k_L X(z)+k_R X_R(\bar z)))$~\footnote{Strictly speaking a
  cocyle factor should be included, see
  \cite[chap. 8.2]{Polchinski:1998rq}; here, just like in conventional
  string theory, this only affects a possible global sign in certain
  amplitudes.}:
\begin{multline}
  \label{eq:KN-def}
    \langle \prod_{j=1}^{n}
    e^{i (k_{Lj} X_L(z_j)+k_{Rj}X_R(\bar z_j))}
    \rangle =\\
     e^{-\sum_{i,j} k_{Li}\cdot k_{Lj} \langle X_L(z_i) X_L(z_j)\rangle}
     e^{-\sum_{i,j} k_{Ri}\cdot k_{Rj} \langle X_R(\bar
       z_i) X_R(\bar z_j)\rangle}
  \end{multline}

  In ordinary string theory, this reduces to the standard expression
  $\prod_{i<j} |z_{ij}|^{\alpha' k_i \cdot k_j}$ or
  $\prod_{i<j} (z_{ij})^{\alpha' k_{Li} \cdot k_{Lj}/2}(\bar
  z_{ij})^{\alpha' k_{Ri} \cdot k_{Rj}/2}$ when windings are included.

  In the twisted
  string, we use \eqref{eq:XrXr} to obtain
\begin{equation}
  \label{eq:KN-twist}
  \prod_{i<j}\bigg((z_i-z_j)^{\frac12
    \alpha' k_{Li} \cdot k_{Lj}}   \times (\bar z_i-\bar z_j)^{-\frac12 \alpha' k_{Ri} \cdot k_{Rj}}   \bigg)\,.
\end{equation}
Given the definitions of the momenta in
\eqref{eq:mom-def}, we have that
$\alpha' k_{L1}\cdot
k_{L2} =\alpha' k_{R1}\cdot k_{R2} -2 n_1 w_2 - 2n_2 w_1 $.
In the presence of windings, eq.~\eqref{eq:KN-twist} is then rewritten
\begin{equation}
  \label{eq:KN-twist-explicit}
  \prod_{i<j} \left(\frac{z_i-z_j}{\bar z_{i}-\bar z_j}\right)^{\frac 12\alpha' q_i\cdot q_j} |z_i-z_j|^{n_i w_j +
    n_j w_i}
\end{equation}
where we used an effective higher-dimensional shorthand notation
$q=(k^m, \frac{n}{R},\frac{wR}{\alpha'})$, such that
$2q_i\cdot q_j = k_{Li}\cdot k_{Lj} + k_{Ri}\cdot k_{Rj} =
2k_i^mk_{jm} +n_in_j/R^2+w_iw_j(R/\alpha')^2$.

To compute these amplitudes we generalize the original
observation of \cite{Huang:2016bdd} which implies a remarkable property: The result of these integrals are rational functions of the
kinematic invariants.
This can be seen explicitly from the following formula:
\begin{multline}
   \int d^2z    \left(\frac z{\bar z}\right)^a | z|^{2n} \left(\frac{1-z}{1-\bar z}\right)^{b} \!\!|1-\bar z|^{2m}
  =\\ \frac{(-a-n)^{\underline{2n+1}}(-b-m)^{\underline{2m+1}}}{(-1-a-b-n-m)^{\underline{2n+2m+3}}}
\label{eq:formula-2}
\end{multline}
where $(a)^{\underline{n}}=\frac{\Gamma(a+n)}{\Gamma(a)}$ is known as a Pochhammer
symbol. It simplifies to a product when $n$ is integer
$(a)^{\underline{n}}= a (a+1)\dots(a+n-1)$.

This rational result is typical of a
field theoretic behaviour at tree-level. By factorisation, higher point
amplitudes should also be given by rational functions.

This surprising fact which starkly contrasts with string theory calculations, where the oscillator spectrum gives rise to infinitely many
poles in the S-matrix, can be understood as follows. The discrete
momenta in physical states require compensating oscillator excitations
(see again eq.~\eqref{eq:Vir-S1}). Momentum quantisation and
conservation in the extra-dimensions therefore imply that only
finitely many states can be exchanged in a given channel~\footnote{
  Not also that these integrals have been 
recently  described in the Kawai-Lewellen-Tye~\cite{Kawai:1985xq} context as twisted
cohomology cycles \cite{Mizera:2017cqs} which, in particular cases, localize to the scattering equations \cite{matsumoto1998}. Properly understood this would give a rigorous foundation for the prescription given in \cite{Siegel:2015axg} to obtain CHY-like formulas.}.

At loop orders, unitarity cuts predicts that the twisted-string
integrand in a loop-momentum formulation~\cite{DHoker:1988pdl} must
also be a rational function.  In ordinary string theory, these integrands
are typically given by elliptic multiple zeta
values~\cite{Broedel:2014vla,DHoker:2015wxz,DHoker:2015gmr,Broedel:2017jdo}
and the fact that these new integrands should be rational is an interesting mathematical fact deserving of further investigation.

\paragraph{CHY integrands for higher spins:}
A prescription for obtaining the scattering equations from the twisted
string was given in \cite{Siegel:2015axg}. We apply it here to obtain
scattering equations for states carrying winding modes and
corresponding Cachazo-He-Yuan~\cite{Cachazo:2013hca} formulae. The
integrand for a generic amplitude in a space with one compact
dimension has a universal contribution from the twisted Koba-Nielsen
factor \eqref{eq:KN-twist}.  Following~\cite{Siegel:2015axg}, we shift
the anti-holomorphic coordinate as
$\bar{z}\rightarrow z -\frac{1}{\beta}\bar{z}$ and take the limit when
$\beta\rightarrow \infty$ so that $\bar z\to z$. Contrary to the
uncompactified case, the exponentials in the Koba-Nielsen factor do
not cancel completely leaving a $(z_{ij})^{n_iw_j+n_jw_i}$
contribution to the amplitude, as can be seen from
eq.~\eqref{eq:KN-twist-explicit}. The scattering equations come at the
next order by integrating over $\bar{z}$ to obtain $\bar{\delta}(E_i)$
where
$E_i=\sum_{i<k}\frac{q_i\cdot
  q_j}{z_{ij}}-\frac{2}{\alpha'}\frac{(n_iw_j+n_jw_i)}{z_{ij}}=0$. In
the limit eq.~\eqref{eq:KN-twist-explicit} becomes
\begin{align}
  \int (\cdots)\prod_{i<j} (z_i-z_j)^{n_iw_j+n_jw_i}\bar{\delta}(E_i).
  \label{eq:general-winding}
\end{align}
The integral supported on the scattering equations, done naively, does not reproduce
the one obtained using formulas like~\eqref{eq:formula-2} (but it still is a
rational function). This is similar to the procedure described in
\cite{Bjerrum-Bohr:2014qwa} where, after changing the integration
measure to one containing the scattering equation in a string
amplitude, one does not have the same amplitude anymore and a limit must be taken.

In our case here, the proper CHY amplitude is obtained after the
$\alpha'\rightarrow\infty$ limit is taken. This decouples the windings
from the scattering equations but leave a contribution of
$(z_i-z_j)^{n_iw_j+n_jw_i} $ to the integrand. We hope this formula paves the way to describe higher spin
(and perhaps massive) amplitudes in the CHY framework.

\paragraph{Partition function:}
The partition function of the bosonic twisted model is readily written
in term of a state sum. The twisted oscillator part was computed in
\cite[below eq.~5.15]{Lee:2017utr}. The only new ingredient here is
the (ordinary) lattice sum coming from the left and right-moving
zero modes. As emphasised above, this piece is unaffected by the changes in
the oscillator sector so the whole partition function is given by
\begin{equation}
  \label{eq:part-fun-winding}
  Z(\tau,\tilde \tau) = \int{d^{26}k}\,
  e^{i\pi (\tau
    -\tilde \tau )\frac{\alpha' k^2}2}\!\!\!\! \sum _{n,m\in \mathbb{Z}^2} \frac{q^{p_L^2/2} \tilde
  q^{p_R^2/2}}{\eta(q)^{24} \eta^{24}(1/\tilde q)}
\end{equation}
where $q=\exp(2i \pi \tau)$, $\tilde q =\exp(-2 i \pi \tilde \tau)$
and $\eta(q) = q^{1/24}\prod_{n=1}^\infty(1-q^n)$. As observed in
\cite{Lee:2017utr}, the unusual $1/\tilde q$ makes a naive
interpretation of the integration variable $\tau$ impossible, for
otherwise $\eta(1/\tilde q)$ is not defined. The model should be
complexified, a matter that will be studied somewhere else and
connected with similar problems in the ambitwistor string side
\cite{Casali:2016atr,Yu:2017bpw,Casali:2017zkz}.

Physical states are obtained in the state sum by integrating
$\int_{-1/2}^{1/2}d( \Re e \tau) Z(\tau,\bar \tau)$, as is standard and
was done in the flat case in \cite{Lee:2017utr}. Higher dimensional
generalisations are done just like in string theory, and it would of
course be interesting to see if some compactification manifolds would
produce better behaved theories, without ghosts for instance.

\section{Discussion}
\label{sec:discussion}
In this letter we described a surprising new facet of twisted strings:
the winding sector. Its existence calls for a complete reevaluation of
the kind of target space theories these models describe and opens the
way to many new interesting possibilities. Many baffling aspects of
this theory need clarification, for instance, what is the 
target space theory describing interacting massless higher spins at the self-dual radius. The
presence of negative-norm states seem to ruin the unitarity of the
theory, but perhaps there is a way to project these states out.

The most urgent question is probably a detailed study of the
$\alpha'\to\infty$ limit, which is expected to generalise the
ambitwistor string framework. In this limit winding modes are
admissible but it is unclear if they contribute or decouple from the
spectrum. This question is under investigation and will be discussed
elsewhere. Closely related is the possibility of introducing consistent
Yang-Mills interactions in the ambitwistor string by taking the
tensionless limit from this model at the self-dual radius and a
possible inclusion of windings in loop amplitudes in the ambitwistor
string~\cite{Adamo:2013tsa,Ohmori:2015sha,Geyer:2015bja}.

Another interesting direction is to use the above formalism to
obtain CHY formulae for higher spin particles, massive particles and
higher dimensional operators generalising the work of \cite{He:2016iqi}.

Finally, the connection with double field theory (DFT) remains to be
elucidated. It would be interesting to compare the amplitude
calculations of these models with the ones in DFT of
\cite{Boels:2015daa} for instance.
On a more conceptual level, the sign change seems to be connected to the
complexification of space-time and exchanging the roles of windings
and momenta. It would be interesting to make a connection with
\cite{Witten:1988zd,Tseytlin:1990va,Green:1991et,Mende:1994wf}. Some
comments on DFT background were also recently made in
\cite{Morand:2017fnv} in relation to Siegel's chiral string (which we
call twisted strings) and it would be interesting to try
to include compact directions in their analysis.

\onecolumngrid

\medskip

\section{Acknowledgements}
We would like to thank David Skinner, Tim Adamo for early discussions on
the existence of winding modes in the ambitwistor string, and David
Berman on their relation to double field theory. We'd also like to
thank Guillaume Bossard, Michael Green, and the above for discussions and
comments on various ideas related to this project. We are also
grateful to David Berman, Michael Green, Alexander Ochirov and Pierre Vanhove for
insightful comments on this manuscript. This
research was supported by the Munich Institute for Astro- and Particle
Physics (MIAPP) of the DFG cluster of excellence ``Origin and
Structure of the Universe''. The work of ED is supported by the EPSRC grant EP/M018911/1, the work of PT is supported by STFC grant ST/L000385/1.

\medskip

\twocolumngrid   


\begin{thebibliography}{50}%
\makeatletter
\providecommand \@ifxundefined [1]{%
 \@ifx{#1\undefined}
}%
\providecommand \@ifnum [1]{%
 \ifnum #1\expandafter \@firstoftwo
 \else \expandafter \@secondoftwo
 \fi
}%
\providecommand \@ifx [1]{%
 \ifx #1\expandafter \@firstoftwo
 \else \expandafter \@secondoftwo
 \fi
}%
\providecommand \natexlab [1]{#1}%
\providecommand \enquote  [1]{``#1''}%
\providecommand \bibnamefont  [1]{#1}%
\providecommand \bibfnamefont [1]{#1}%
\providecommand \citenamefont [1]{#1}%
\providecommand \href@noop [0]{\@secondoftwo}%
\providecommand \href [0]{\begingroup \@sanitize@url \@href}%
\providecommand \@href[1]{\@@startlink{#1}\@@href}%
\providecommand \@@href[1]{\endgroup#1\@@endlink}%
\providecommand \@sanitize@url [0]{\catcode `\\12\catcode `\$12\catcode
  `\&12\catcode `\#12\catcode `\^12\catcode `\_12\catcode `\%12\relax}%
\providecommand \@@startlink[1]{}%
\providecommand \@@endlink[0]{}%
\providecommand \url  [0]{\begingroup\@sanitize@url \@url }%
\providecommand \@url [1]{\endgroup\@href {#1}{\urlprefix }}%
\providecommand \urlprefix  [0]{URL }%
\providecommand \Eprint [0]{\href }%
\providecommand \doibase [0]{http://dx.doi.org/}%
\providecommand \selectlanguage [0]{\@gobble}%
\providecommand \bibinfo  [0]{\@secondoftwo}%
\providecommand \bibfield  [0]{\@secondoftwo}%
\providecommand \translation [1]{[#1]}%
\providecommand \BibitemOpen [0]{}%
\providecommand \bibitemStop [0]{}%
\providecommand \bibitemNoStop [0]{.\EOS\space}%
\providecommand \EOS [0]{\spacefactor3000\relax}%
\providecommand \BibitemShut  [1]{\csname bibitem#1\endcsname}%
\let\auto@bib@innerbib\@empty
\bibitem [{\citenamefont {Witten}(2004)}]{Witten:2003nn}%
  \BibitemOpen
  \bibfield  {author} {\bibinfo {author} {\bibfnamefont {E.}~\bibnamefont
  {Witten}},\ }\href {\doibase 10.1007/s00220-004-1187-3} {\bibfield  {journal}
  {\bibinfo  {journal} {Commun. Math. Phys.}\ }\textbf {\bibinfo {volume}
  {252}},\ \bibinfo {pages} {189} (\bibinfo {year} {2004})},\ \Eprint
  {http://arxiv.org/abs/hep-th/0312171} {arXiv:hep-th/0312171 [hep-th]}
  \BibitemShut {NoStop}%
\bibitem [{\citenamefont {Cachazo}\ \emph
  {et~al.}(2014{\natexlab{a}})\citenamefont {Cachazo}, \citenamefont {He},\
  and\ \citenamefont {Yuan}}]{Cachazo:2013iea}%
  \BibitemOpen
  \bibfield  {author} {\bibinfo {author} {\bibfnamefont {F.}~\bibnamefont
  {Cachazo}}, \bibinfo {author} {\bibfnamefont {S.}~\bibnamefont {He}}, \ and\
  \bibinfo {author} {\bibfnamefont {E.~Y.}\ \bibnamefont {Yuan}},\ }\href
  {\doibase 10.1007/JHEP07(2014)033} {\bibfield  {journal} {\bibinfo  {journal}
  {JHEP}\ }\textbf {\bibinfo {volume} {1407}},\ \bibinfo {pages} {033}
  (\bibinfo {year} {2014}{\natexlab{a}})},\ \Eprint
  {http://arxiv.org/abs/1309.0885} {arXiv:1309.0885 [hep-th]} \BibitemShut
  {NoStop}%
\bibitem [{\citenamefont {Cachazo}\ \emph
  {et~al.}(2014{\natexlab{b}})\citenamefont {Cachazo}, \citenamefont {He},\
  and\ \citenamefont {Yuan}}]{Cachazo:2013hca}%
  \BibitemOpen
  \bibfield  {author} {\bibinfo {author} {\bibfnamefont {F.}~\bibnamefont
  {Cachazo}}, \bibinfo {author} {\bibfnamefont {S.}~\bibnamefont {He}}, \ and\
  \bibinfo {author} {\bibfnamefont {E.~Y.}\ \bibnamefont {Yuan}},\ }\href
  {\doibase 10.1103/PhysRevLett.113.171601} {\bibfield  {journal} {\bibinfo
  {journal} {Phys.Rev.Lett.}\ }\textbf {\bibinfo {volume} {113}},\ \bibinfo
  {pages} {171601} (\bibinfo {year} {2014}{\natexlab{b}})},\ \Eprint
  {http://arxiv.org/abs/1307.2199} {arXiv:1307.2199 [hep-th]} \BibitemShut
  {NoStop}%
\bibitem [{\citenamefont {Cachazo}\ \emph
  {et~al.}(2014{\natexlab{c}})\citenamefont {Cachazo}, \citenamefont {He},\
  and\ \citenamefont {Yuan}}]{Cachazo:2013gna}%
  \BibitemOpen
  \bibfield  {author} {\bibinfo {author} {\bibfnamefont {F.}~\bibnamefont
  {Cachazo}}, \bibinfo {author} {\bibfnamefont {S.}~\bibnamefont {He}}, \ and\
  \bibinfo {author} {\bibfnamefont {E.~Y.}\ \bibnamefont {Yuan}},\ }\href
  {\doibase 10.1103/PhysRevD.90.065001} {\bibfield  {journal} {\bibinfo
  {journal} {Phys. Rev.}\ }\textbf {\bibinfo {volume} {D90}},\ \bibinfo {pages}
  {065001} (\bibinfo {year} {2014}{\natexlab{c}})},\ \Eprint
  {http://arxiv.org/abs/1306.6575} {arXiv:1306.6575 [hep-th]} \BibitemShut
  {NoStop}%
\bibitem [{\citenamefont {Adamo}\ \emph {et~al.}(2014)\citenamefont {Adamo},
  \citenamefont {Casali},\ and\ \citenamefont {Skinner}}]{Adamo:2013tsa}%
  \BibitemOpen
  \bibfield  {author} {\bibinfo {author} {\bibfnamefont {T.}~\bibnamefont
  {Adamo}}, \bibinfo {author} {\bibfnamefont {E.}~\bibnamefont {Casali}}, \
  and\ \bibinfo {author} {\bibfnamefont {D.}~\bibnamefont {Skinner}},\ }\href
  {\doibase 10.1007/JHEP04(2014)104} {\bibfield  {journal} {\bibinfo  {journal}
  {JHEP}\ }\textbf {\bibinfo {volume} {04}},\ \bibinfo {pages} {104} (\bibinfo
  {year} {2014})},\ \Eprint {http://arxiv.org/abs/1312.3828} {arXiv:1312.3828
  [hep-th]} \BibitemShut {NoStop}%
\bibitem [{\citenamefont {Adamo}\ and\ \citenamefont
  {Casali}(2015)}]{Adamo:2015hoa}%
  \BibitemOpen
  \bibfield  {author} {\bibinfo {author} {\bibfnamefont {T.}~\bibnamefont
  {Adamo}}\ and\ \bibinfo {author} {\bibfnamefont {E.}~\bibnamefont {Casali}},\
  }\href {\doibase 10.1007/JHEP05(2015)120} {\bibfield  {journal} {\bibinfo
  {journal} {JHEP}\ }\textbf {\bibinfo {volume} {05}},\ \bibinfo {pages} {120}
  (\bibinfo {year} {2015})},\ \Eprint {http://arxiv.org/abs/1502.06826}
  {arXiv:1502.06826 [hep-th]} \BibitemShut {NoStop}%
\bibitem [{\citenamefont {Ohmori}(2015)}]{Ohmori:2015sha}%
  \BibitemOpen
  \bibfield  {author} {\bibinfo {author} {\bibfnamefont {K.}~\bibnamefont
  {Ohmori}},\ }\href {\doibase 10.1007/JHEP06(2015)075} {\bibfield  {journal}
  {\bibinfo  {journal} {JHEP}\ }\textbf {\bibinfo {volume} {06}},\ \bibinfo
  {pages} {075} (\bibinfo {year} {2015})},\ \Eprint
  {http://arxiv.org/abs/1504.02675} {arXiv:1504.02675 [hep-th]} \BibitemShut
  {NoStop}%
\bibitem [{\citenamefont {Geyer}\ \emph {et~al.}(2015)\citenamefont {Geyer},
  \citenamefont {Mason}, \citenamefont {Monteiro},\ and\ \citenamefont
  {Tourkine}}]{Geyer:2015bja}%
  \BibitemOpen
  \bibfield  {author} {\bibinfo {author} {\bibfnamefont {Y.}~\bibnamefont
  {Geyer}}, \bibinfo {author} {\bibfnamefont {L.}~\bibnamefont {Mason}},
  \bibinfo {author} {\bibfnamefont {R.}~\bibnamefont {Monteiro}}, \ and\
  \bibinfo {author} {\bibfnamefont {P.}~\bibnamefont {Tourkine}},\ }\href
  {\doibase 10.1103/PhysRevLett.115.121603} {\bibfield  {journal} {\bibinfo
  {journal} {Phys. Rev. Lett.}\ }\textbf {\bibinfo {volume} {115}},\ \bibinfo
  {pages} {121603} (\bibinfo {year} {2015})},\ \Eprint
  {http://arxiv.org/abs/1507.00321} {arXiv:1507.00321 [hep-th]} \BibitemShut
  {NoStop}%
\bibitem [{\citenamefont {Geyer}\ \emph {et~al.}(2016)\citenamefont {Geyer},
  \citenamefont {Mason}, \citenamefont {Monteiro},\ and\ \citenamefont
  {Tourkine}}]{Geyer:2015jch}%
  \BibitemOpen
  \bibfield  {author} {\bibinfo {author} {\bibfnamefont {Y.}~\bibnamefont
  {Geyer}}, \bibinfo {author} {\bibfnamefont {L.}~\bibnamefont {Mason}},
  \bibinfo {author} {\bibfnamefont {R.}~\bibnamefont {Monteiro}}, \ and\
  \bibinfo {author} {\bibfnamefont {P.}~\bibnamefont {Tourkine}},\ }\href
  {\doibase 10.1007/JHEP03(2016)114} {\bibfield  {journal} {\bibinfo  {journal}
  {JHEP}\ }\textbf {\bibinfo {volume} {03}},\ \bibinfo {pages} {114} (\bibinfo
  {year} {2016})},\ \Eprint {http://arxiv.org/abs/1511.06315} {arXiv:1511.06315
  [hep-th]} \BibitemShut {NoStop}%
\bibitem [{\citenamefont {Adamo}\ \emph {et~al.}(2015)\citenamefont {Adamo},
  \citenamefont {Casali},\ and\ \citenamefont {Skinner}}]{Adamo:2014wea}%
  \BibitemOpen
  \bibfield  {author} {\bibinfo {author} {\bibfnamefont {T.}~\bibnamefont
  {Adamo}}, \bibinfo {author} {\bibfnamefont {E.}~\bibnamefont {Casali}}, \
  and\ \bibinfo {author} {\bibfnamefont {D.}~\bibnamefont {Skinner}},\ }\href
  {\doibase 10.1007/JHEP02(2015)116} {\bibfield  {journal} {\bibinfo  {journal}
  {JHEP}\ }\textbf {\bibinfo {volume} {02}},\ \bibinfo {pages} {116} (\bibinfo
  {year} {2015})},\ \Eprint {http://arxiv.org/abs/1409.5656} {arXiv:1409.5656
  [hep-th]} \BibitemShut {NoStop}%
\bibitem [{\citenamefont {Adamo}\ \emph {et~al.}(2017)\citenamefont {Adamo},
  \citenamefont {Casali}, \citenamefont {Mason},\ and\ \citenamefont
  {Nekovar}}]{Adamo:2017sze}%
  \BibitemOpen
  \bibfield  {author} {\bibinfo {author} {\bibfnamefont {T.}~\bibnamefont
  {Adamo}}, \bibinfo {author} {\bibfnamefont {E.}~\bibnamefont {Casali}},
  \bibinfo {author} {\bibfnamefont {L.}~\bibnamefont {Mason}}, \ and\ \bibinfo
  {author} {\bibfnamefont {S.}~\bibnamefont {Nekovar}},\ }\href {\doibase
  10.1007/JHEP11(2017)160} {\bibfield  {journal} {\bibinfo  {journal} {JHEP}\
  }\textbf {\bibinfo {volume} {11}},\ \bibinfo {pages} {160} (\bibinfo {year}
  {2017})},\ \Eprint {http://arxiv.org/abs/1708.09249} {arXiv:1708.09249
  [hep-th]} \BibitemShut {NoStop}%
\bibitem [{\citenamefont {Siegel}(2015)}]{Siegel:2015axg}%
  \BibitemOpen
  \bibfield  {author} {\bibinfo {author} {\bibfnamefont {W.}~\bibnamefont
  {Siegel}},\ }\href@noop {} {\  (\bibinfo {year} {2015})},\ \Eprint
  {http://arxiv.org/abs/1512.02569} {arXiv:1512.02569 [hep-th]} \BibitemShut
  {NoStop}%
\bibitem [{\citenamefont {Casali}\ and\ \citenamefont
  {Tourkine}(2016)}]{Casali:2016atr}%
  \BibitemOpen
  \bibfield  {author} {\bibinfo {author} {\bibfnamefont {E.}~\bibnamefont
  {Casali}}\ and\ \bibinfo {author} {\bibfnamefont {P.}~\bibnamefont
  {Tourkine}},\ }\href@noop {} {\  (\bibinfo {year} {2016})},\ \Eprint
  {http://arxiv.org/abs/1606.05636} {arXiv:1606.05636 [hep-th]} \BibitemShut
  {NoStop}%
\bibitem [{\citenamefont {Lee}\ \emph {et~al.}(2017)\citenamefont {Lee},
  \citenamefont {Rey},\ and\ \citenamefont {Rosabal}}]{Lee:2017utr}%
  \BibitemOpen
  \bibfield  {author} {\bibinfo {author} {\bibfnamefont {K.}~\bibnamefont
  {Lee}}, \bibinfo {author} {\bibfnamefont {S.-J.}\ \bibnamefont {Rey}}, \ and\
  \bibinfo {author} {\bibfnamefont {J.~A.}\ \bibnamefont {Rosabal}},\
  }\href@noop {} {\  (\bibinfo {year} {2017})},\ \Eprint
  {http://arxiv.org/abs/1708.05707} {arXiv:1708.05707 [hep-th]} \BibitemShut
  {NoStop}%
\bibitem [{\citenamefont {Azevedo}\ and\ \citenamefont
  {Jusinskas}(2017)}]{Azevedo:2017yjy}%
  \BibitemOpen
  \bibfield  {author} {\bibinfo {author} {\bibfnamefont {T.}~\bibnamefont
  {Azevedo}}\ and\ \bibinfo {author} {\bibfnamefont {R.~L.}\ \bibnamefont
  {Jusinskas}},\ }\href@noop {} {\  (\bibinfo {year} {2017})},\ \Eprint
  {http://arxiv.org/abs/1707.08840} {arXiv:1707.08840 [hep-th]} \BibitemShut
  {NoStop}%
\bibitem [{Note1()}]{Note1}%
  \BibitemOpen
  \bibinfo {note} {This is counter-intuitive because field theories are also
  found in the infinite tension limit of conventional string theory. Note also
  that we choose the terminology ``twisted'' instead of ``chiral'' or
  ``left-handed'' as first used in~\cite {Huang:2016bdd,Siegel:2015axg} because
  our string is not chiral.}\BibitemShut {Stop}%
\bibitem [{\citenamefont {Amati}\ \emph {et~al.}(1987)\citenamefont {Amati},
  \citenamefont {Ciafaloni},\ and\ \citenamefont {Veneziano}}]{Amati:1987wq}%
  \BibitemOpen
  \bibfield  {author} {\bibinfo {author} {\bibfnamefont {D.}~\bibnamefont
  {Amati}}, \bibinfo {author} {\bibfnamefont {M.}~\bibnamefont {Ciafaloni}}, \
  and\ \bibinfo {author} {\bibfnamefont {G.}~\bibnamefont {Veneziano}},\ }\href
  {\doibase 10.1016/0370-2693(87)90346-7} {\bibfield  {journal} {\bibinfo
  {journal} {Phys. Lett.}\ }\textbf {\bibinfo {volume} {B197}},\ \bibinfo
  {pages} {81} (\bibinfo {year} {1987})}\BibitemShut {NoStop}%
\bibitem [{\citenamefont {Gross}\ and\ \citenamefont
  {Mende}(1988)}]{Gross:1987ar}%
  \BibitemOpen
  \bibfield  {author} {\bibinfo {author} {\bibfnamefont {D.~J.}\ \bibnamefont
  {Gross}}\ and\ \bibinfo {author} {\bibfnamefont {P.~F.}\ \bibnamefont
  {Mende}},\ }\href {\doibase 10.1016/0550-3213(88)90390-2} {\bibfield
  {journal} {\bibinfo  {journal} {Nucl. Phys.}\ }\textbf {\bibinfo {volume}
  {B303}},\ \bibinfo {pages} {407} (\bibinfo {year} {1988})}\BibitemShut
  {NoStop}%
\bibitem [{\citenamefont {Mason}\ and\ \citenamefont
  {Skinner}(2014)}]{Mason:2013sva}%
  \BibitemOpen
  \bibfield  {author} {\bibinfo {author} {\bibfnamefont {L.}~\bibnamefont
  {Mason}}\ and\ \bibinfo {author} {\bibfnamefont {D.}~\bibnamefont
  {Skinner}},\ }\href {\doibase 10.1007/JHEP07(2014)048} {\bibfield  {journal}
  {\bibinfo  {journal} {JHEP}\ }\textbf {\bibinfo {volume} {07}},\ \bibinfo
  {pages} {048} (\bibinfo {year} {2014})},\ \Eprint
  {http://arxiv.org/abs/1311.2564} {arXiv:1311.2564 [hep-th]} \BibitemShut
  {NoStop}%
\bibitem [{Note2()}]{Note2}%
  \BibitemOpen
  \bibinfo {note} {See also the older work \cite {Hwang:1998gs}.}\BibitemShut
  {Stop}%
\bibitem [{\citenamefont {Bagchi}\ \emph {et~al.}(2016)\citenamefont {Bagchi},
  \citenamefont {Chakrabortty},\ and\ \citenamefont {Parekh}}]{Bagchi:2015nca}%
  \BibitemOpen
  \bibfield  {author} {\bibinfo {author} {\bibfnamefont {A.}~\bibnamefont
  {Bagchi}}, \bibinfo {author} {\bibfnamefont {S.}~\bibnamefont
  {Chakrabortty}}, \ and\ \bibinfo {author} {\bibfnamefont {P.}~\bibnamefont
  {Parekh}},\ }\href {\doibase 10.1007/JHEP01(2016)158} {\bibfield  {journal}
  {\bibinfo  {journal} {JHEP}\ }\textbf {\bibinfo {volume} {01}},\ \bibinfo
  {pages} {158} (\bibinfo {year} {2016})},\ \Eprint
  {http://arxiv.org/abs/1507.04361} {arXiv:1507.04361 [hep-th]} \BibitemShut
  {NoStop}%
\bibitem [{\citenamefont {Hwang}\ \emph {et~al.}(1999)\citenamefont {Hwang},
  \citenamefont {Marnelius},\ and\ \citenamefont {Saltsidis}}]{Hwang:1998gs}%
  \BibitemOpen
  \bibfield  {author} {\bibinfo {author} {\bibfnamefont {S.}~\bibnamefont
  {Hwang}}, \bibinfo {author} {\bibfnamefont {R.}~\bibnamefont {Marnelius}}, \
  and\ \bibinfo {author} {\bibfnamefont {P.}~\bibnamefont {Saltsidis}},\ }\href
  {\doibase 10.1063/1.532994} {\bibfield  {journal} {\bibinfo  {journal} {J.
  Math. Phys.}\ }\textbf {\bibinfo {volume} {40}},\ \bibinfo {pages} {4639}
  (\bibinfo {year} {1999})},\ \Eprint {http://arxiv.org/abs/hep-th/9804003}
  {arXiv:hep-th/9804003 [hep-th]} \BibitemShut {NoStop}%
\bibitem [{Note3()}]{Note3}%
  \BibitemOpen
  \bibinfo {note} {See \cite {Lee:2017utr} for more details on these
  expressions.}\BibitemShut {Stop}%
\bibitem [{\citenamefont {Leite}\ and\ \citenamefont
  {Siegel}(2016)}]{Leite:2016fno}%
  \BibitemOpen
  \bibfield  {author} {\bibinfo {author} {\bibfnamefont {M.~M.}\ \bibnamefont
  {Leite}}\ and\ \bibinfo {author} {\bibfnamefont {W.}~\bibnamefont {Siegel}},\
  }\href@noop {} {\  (\bibinfo {year} {2016})},\ \Eprint
  {http://arxiv.org/abs/1610.02052} {arXiv:1610.02052 [hep-th]} \BibitemShut
  {NoStop}%
\bibitem [{\citenamefont {Polchinski}(2007{\natexlab{a}})}]{Polchinski:1998rr}%
  \BibitemOpen
  \bibfield  {author} {\bibinfo {author} {\bibfnamefont {J.}~\bibnamefont
  {Polchinski}},\ }\href@noop {} {\emph {\bibinfo {title} {{String Theory. Vol.
  2: Superstring Theory and Beyond}}}}\ (\bibinfo  {publisher} {Cambridge
  University Press},\ \bibinfo {year} {2007})\BibitemShut {NoStop}%
\bibitem [{Note4()}]{Note4}%
  \BibitemOpen
  \bibinfo {note} {
  $\bar L_{-m}|\mathrm{phys}\rangle=0$ with $m>0$, there still remain
  infinitely many physical negative-norm states, as can be easily
  checked.
    }\BibitemShut
  {NoStop}%
\bibitem [{\citenamefont {Buscher}(1988)}]{Buscher:1987qj}%
  \BibitemOpen
  \bibfield  {author} {\bibinfo {author} {\bibfnamefont {T.~H.}\ \bibnamefont
  {Buscher}},\ }\href {\doibase 10.1016/0370-2693(88)90602-8} {\bibfield
  {journal} {\bibinfo  {journal} {Phys. Lett.}\ }\textbf {\bibinfo {volume}
  {B201}},\ \bibinfo {pages} {466} (\bibinfo {year} {1988})}\BibitemShut
  {NoStop}%
\bibitem [{Note5()}]{Note5}%
  \BibitemOpen
  \bibinfo {note} {Note further that for rational values of $R/\protect \sqrt
  {\alpha '}=p/q\in \protect \mathbb {Q}$ infinitely many massless states are
  generated. This is more evidence of the unconventionality of the target space
  theory.}\BibitemShut {Stop}%
\bibitem [{\citenamefont {Huang}\ \emph {et~al.}(2016)\citenamefont {Huang},
  \citenamefont {Siegel},\ and\ \citenamefont {Yuan}}]{Huang:2016bdd}%
  \BibitemOpen
  \bibfield  {author} {\bibinfo {author} {\bibfnamefont {Y.-t.}\ \bibnamefont
  {Huang}}, \bibinfo {author} {\bibfnamefont {W.}~\bibnamefont {Siegel}}, \
  and\ \bibinfo {author} {\bibfnamefont {E.~Y.}\ \bibnamefont {Yuan}},\
  }\href@noop {} {\  (\bibinfo {year} {2016})},\ \Eprint
  {http://arxiv.org/abs/1603.02588} {arXiv:1603.02588 [hep-th]} \BibitemShut
  {NoStop}%
\bibitem [{\citenamefont {Tseytlin}(1991)}]{Tseytlin:1990va}%
  \BibitemOpen
  \bibfield  {author} {\bibinfo {author} {\bibfnamefont {A.~A.}\ \bibnamefont
  {Tseytlin}},\ }\href {\doibase 10.1016/0550-3213(91)90266-Z} {\bibfield
  {journal} {\bibinfo  {journal} {Nucl. Phys.}\ }\textbf {\bibinfo {volume}
  {B350}},\ \bibinfo {pages} {395} (\bibinfo {year} {1991})}\BibitemShut
  {NoStop}%
\bibitem [{Note6()}]{Note6}%
  \BibitemOpen
  \bibinfo {note} {Strictly speaking a cocyle factor should be included, see
  \cite [chap. 8.2]{Polchinski:1998rq}; here, just like in conventional string
  theory, this only affects a possible global sign in certain
  amplitudes.}\BibitemShut {Stop}%
\bibitem [{Note7()}]{Note7}%
  \BibitemOpen
  \bibinfo {note} {Not also that these integrals have been recently described
  in the Kawai-Lewellen-Tye~\cite {Kawai:1985xq} context as twisted cohomology
  cycles \cite {Mizera:2017cqs} which, in particular cases, localize to the
  scattering equations \cite {matsumoto1998}. Properly understood this would
  give a rigorous foundation for the prescription given in \cite
  {Siegel:2015axg} to obtain CHY-like formulas.}\BibitemShut {Stop}%
\bibitem [{\citenamefont {D'Hoker}\ and\ \citenamefont
  {Phong}(1988)}]{DHoker:1988pdl}%
  \BibitemOpen
  \bibfield  {author} {\bibinfo {author} {\bibfnamefont {E.}~\bibnamefont
  {D'Hoker}}\ and\ \bibinfo {author} {\bibfnamefont {D.~H.}\ \bibnamefont
  {Phong}},\ }\href {\doibase 10.1103/RevModPhys.60.917} {\bibfield  {journal}
  {\bibinfo  {journal} {Rev. Mod. Phys.}\ }\textbf {\bibinfo {volume} {60}},\
  \bibinfo {pages} {917} (\bibinfo {year} {1988})}\BibitemShut {NoStop}%
\bibitem [{\citenamefont {Broedel}\ \emph {et~al.}(2015)\citenamefont
  {Broedel}, \citenamefont {Mafra}, \citenamefont {Matthes},\ and\
  \citenamefont {Schlotterer}}]{Broedel:2014vla}%
  \BibitemOpen
  \bibfield  {author} {\bibinfo {author} {\bibfnamefont {J.}~\bibnamefont
  {Broedel}}, \bibinfo {author} {\bibfnamefont {C.~R.}\ \bibnamefont {Mafra}},
  \bibinfo {author} {\bibfnamefont {N.}~\bibnamefont {Matthes}}, \ and\
  \bibinfo {author} {\bibfnamefont {O.}~\bibnamefont {Schlotterer}},\ }\href
  {\doibase 10.1007/JHEP07(2015)112} {\bibfield  {journal} {\bibinfo  {journal}
  {JHEP}\ }\textbf {\bibinfo {volume} {07}},\ \bibinfo {pages} {112} (\bibinfo
  {year} {2015})},\ \Eprint {http://arxiv.org/abs/1412.5535} {arXiv:1412.5535
  [hep-th]} \BibitemShut {NoStop}%
\bibitem [{\citenamefont {D'Hoker}\ \emph {et~al.}(2017)\citenamefont
  {D'Hoker}, \citenamefont {Green}, \citenamefont {Gurdogan},\ and\
  \citenamefont {Vanhove}}]{DHoker:2015wxz}%
  \BibitemOpen
  \bibfield  {author} {\bibinfo {author} {\bibfnamefont {E.}~\bibnamefont
  {D'Hoker}}, \bibinfo {author} {\bibfnamefont {M.~B.}\ \bibnamefont {Green}},
  \bibinfo {author} {\bibfnamefont {O.}~\bibnamefont {Gurdogan}}, \ and\
  \bibinfo {author} {\bibfnamefont {P.}~\bibnamefont {Vanhove}},\ }\href
  {\doibase 10.4310/CNTP.2017.v11.n1.a4} {\bibfield  {journal} {\bibinfo
  {journal} {Commun. Num. Theor. Phys.}\ }\textbf {\bibinfo {volume} {11}},\
  \bibinfo {pages} {165} (\bibinfo {year} {2017})},\ \Eprint
  {http://arxiv.org/abs/1512.06779} {arXiv:1512.06779 [hep-th]} \BibitemShut
  {NoStop}%
\bibitem [{\citenamefont {D'Hoker}\ \emph {et~al.}(2015)\citenamefont
  {D'Hoker}, \citenamefont {Green},\ and\ \citenamefont
  {Vanhove}}]{DHoker:2015gmr}%
  \BibitemOpen
  \bibfield  {author} {\bibinfo {author} {\bibfnamefont {E.}~\bibnamefont
  {D'Hoker}}, \bibinfo {author} {\bibfnamefont {M.~B.}\ \bibnamefont {Green}},
  \ and\ \bibinfo {author} {\bibfnamefont {P.}~\bibnamefont {Vanhove}},\ }\href
  {\doibase 10.1007/JHEP08(2015)041} {\bibfield  {journal} {\bibinfo  {journal}
  {JHEP}\ }\textbf {\bibinfo {volume} {08}},\ \bibinfo {pages} {041} (\bibinfo
  {year} {2015})},\ \Eprint {http://arxiv.org/abs/1502.06698} {arXiv:1502.06698
  [hep-th]} \BibitemShut {NoStop}%
\bibitem [{\citenamefont {Broedel}\ \emph {et~al.}(2017)\citenamefont
  {Broedel}, \citenamefont {Matthes}, \citenamefont {Richter},\ and\
  \citenamefont {SCHLotterer}}]{Broedel:2017jdo}%
  \BibitemOpen
  \bibfield  {author} {\bibinfo {author} {\bibfnamefont {J.}~\bibnamefont
  {Broedel}}, \bibinfo {author} {\bibfnamefont {N.}~\bibnamefont {Matthes}},
  \bibinfo {author} {\bibfnamefont {G.}~\bibnamefont {Richter}}, \ and\
  \bibinfo {author} {\bibfnamefont {O.}~\bibnamefont {SCHLotterer}},\
  }\href@noop {} {\  (\bibinfo {year} {2017})},\ \Eprint
  {http://arxiv.org/abs/1704.03449} {arXiv:1704.03449 [hep-th]} \BibitemShut
  {NoStop}%
\bibitem [{\citenamefont {Bjerrum-Bohr}\ \emph {et~al.}(2014)\citenamefont
  {Bjerrum-Bohr}, \citenamefont {Damgaard}, \citenamefont {Tourkine},\ and\
  \citenamefont {Vanhove}}]{Bjerrum-Bohr:2014qwa}%
  \BibitemOpen
  \bibfield  {author} {\bibinfo {author} {\bibfnamefont {N.~E.~J.}\
  \bibnamefont {Bjerrum-Bohr}}, \bibinfo {author} {\bibfnamefont {P.~H.}\
  \bibnamefont {Damgaard}}, \bibinfo {author} {\bibfnamefont {P.}~\bibnamefont
  {Tourkine}}, \ and\ \bibinfo {author} {\bibfnamefont {P.}~\bibnamefont
  {Vanhove}},\ }\href {\doibase 10.1103/PhysRevD.90.106002} {\bibfield
  {journal} {\bibinfo  {journal} {Phys. Rev.}\ }\textbf {\bibinfo {volume}
  {D90}},\ \bibinfo {pages} {106002} (\bibinfo {year} {2014})},\ \Eprint
  {http://arxiv.org/abs/1403.4553} {arXiv:1403.4553 [hep-th]} \BibitemShut
  {NoStop}%
\bibitem [{\citenamefont {Yu}\ \emph {et~al.}(2017)\citenamefont {Yu},
  \citenamefont {Zhang},\ and\ \citenamefont {Zhang}}]{Yu:2017bpw}%
  \BibitemOpen
  \bibfield  {author} {\bibinfo {author} {\bibfnamefont {M.}~\bibnamefont
  {Yu}}, \bibinfo {author} {\bibfnamefont {C.}~\bibnamefont {Zhang}}, \ and\
  \bibinfo {author} {\bibfnamefont {Y.-Z.}\ \bibnamefont {Zhang}},\ }\href@noop
  {} {\  (\bibinfo {year} {2017})},\ \Eprint {http://arxiv.org/abs/1704.01290}
  {arXiv:1704.01290 [hep-th]} \BibitemShut {NoStop}%
\bibitem [{\citenamefont {Casali}\ \emph {et~al.}(2017)\citenamefont {Casali},
  \citenamefont {Herfray},\ and\ \citenamefont {Tourkine}}]{Casali:2017zkz}%
  \BibitemOpen
  \bibfield  {author} {\bibinfo {author} {\bibfnamefont {E.}~\bibnamefont
  {Casali}}, \bibinfo {author} {\bibfnamefont {Y.}~\bibnamefont {Herfray}}, \
  and\ \bibinfo {author} {\bibfnamefont {P.}~\bibnamefont {Tourkine}},\
  }\href@noop {} {\  (\bibinfo {year} {2017})},\ \Eprint
  {http://arxiv.org/abs/1707.09900} {arXiv:1707.09900 [hep-th]} \BibitemShut
  {NoStop}%
\bibitem [{\citenamefont {He}\ and\ \citenamefont {Zhang}(2017)}]{He:2016iqi}%
  \BibitemOpen
  \bibfield  {author} {\bibinfo {author} {\bibfnamefont {S.}~\bibnamefont
  {He}}\ and\ \bibinfo {author} {\bibfnamefont {Y.}~\bibnamefont {Zhang}},\
  }\href {\doibase 10.1007/JHEP02(2017)019} {\bibfield  {journal} {\bibinfo
  {journal} {JHEP}\ }\textbf {\bibinfo {volume} {02}},\ \bibinfo {pages} {019}
  (\bibinfo {year} {2017})},\ \Eprint {http://arxiv.org/abs/1608.08448}
  {arXiv:1608.08448 [hep-th]} \BibitemShut {NoStop}%
\bibitem [{\citenamefont {Boels}\ and\ \citenamefont
  {Horst}(2016)}]{Boels:2015daa}%
  \BibitemOpen
  \bibfield  {author} {\bibinfo {author} {\bibfnamefont {R.~H.}\ \bibnamefont
  {Boels}}\ and\ \bibinfo {author} {\bibfnamefont {C.}~\bibnamefont {Horst}},\
  }\href {\doibase 10.1007/JHEP04(2016)120} {\bibfield  {journal} {\bibinfo
  {journal} {JHEP}\ }\textbf {\bibinfo {volume} {04}},\ \bibinfo {pages} {120}
  (\bibinfo {year} {2016})},\ \Eprint {http://arxiv.org/abs/1512.03192}
  {arXiv:1512.03192 [hep-th]} \BibitemShut {NoStop}%
\bibitem [{\citenamefont {Witten}(1988)}]{Witten:1988zd}%
  \BibitemOpen
  \bibfield  {author} {\bibinfo {author} {\bibfnamefont {E.}~\bibnamefont
  {Witten}},\ }\href {\doibase 10.1103/PhysRevLett.61.670} {\bibfield
  {journal} {\bibinfo  {journal} {Phys. Rev. Lett.}\ }\textbf {\bibinfo
  {volume} {61}},\ \bibinfo {pages} {670} (\bibinfo {year} {1988})}\BibitemShut
  {NoStop}%
\bibitem [{\citenamefont {Green}(1991)}]{Green:1991et}%
  \BibitemOpen
  \bibfield  {author} {\bibinfo {author} {\bibfnamefont {M.~B.}\ \bibnamefont
  {Green}},\ }\href {\doibase 10.1016/0370-2693(91)91048-Z} {\bibfield
  {journal} {\bibinfo  {journal} {Phys. Lett.}\ }\textbf {\bibinfo {volume}
  {B266}},\ \bibinfo {pages} {325} (\bibinfo {year} {1991})}\BibitemShut
  {NoStop}%
\bibitem [{\citenamefont {Mende}(1994)}]{Mende:1994wf}%
  \BibitemOpen
  \bibfield  {author} {\bibinfo {author} {\bibfnamefont {P.~F.}\ \bibnamefont
  {Mende}},\ }\href {\doibase 10.1016/0370-2693(94)91313-7} {\bibfield
  {journal} {\bibinfo  {journal} {Phys. Lett.}\ }\textbf {\bibinfo {volume}
  {B326}},\ \bibinfo {pages} {216} (\bibinfo {year} {1994})},\ \Eprint
  {http://arxiv.org/abs/hep-th/9401126} {arXiv:hep-th/9401126 [hep-th]}
  \BibitemShut {NoStop}%
\bibitem [{\citenamefont {Morand}\ and\ \citenamefont
  {Park}(2017)}]{Morand:2017fnv}%
  \BibitemOpen
  \bibfield  {author} {\bibinfo {author} {\bibfnamefont {K.}~\bibnamefont
  {Morand}}\ and\ \bibinfo {author} {\bibfnamefont {J.-H.}\ \bibnamefont
  {Park}},\ }\href@noop {} {\  (\bibinfo {year} {2017})},\ \Eprint
  {http://arxiv.org/abs/1707.03713} {arXiv:1707.03713 [hep-th]} \BibitemShut
  {NoStop}%
\bibitem [{\citenamefont {Polchinski}(2007{\natexlab{b}})}]{Polchinski:1998rq}%
  \BibitemOpen
  \bibfield  {author} {\bibinfo {author} {\bibfnamefont {J.}~\bibnamefont
  {Polchinski}},\ }\href@noop {} {\emph {\bibinfo {title} {{String Theory. Vol.
  1: an Introduction to the Bosonic String}}}}\ (\bibinfo  {publisher}
  {Cambridge University Press},\ \bibinfo {year} {2007})\BibitemShut {NoStop}%
\bibitem [{\citenamefont {Kawai}\ \emph {et~al.}(1986)\citenamefont {Kawai},
  \citenamefont {Lewellen},\ and\ \citenamefont {Tye}}]{Kawai:1985xq}%
  \BibitemOpen
  \bibfield  {author} {\bibinfo {author} {\bibfnamefont {H.}~\bibnamefont
  {Kawai}}, \bibinfo {author} {\bibfnamefont {D.~C.}\ \bibnamefont {Lewellen}},
  \ and\ \bibinfo {author} {\bibfnamefont {S.~H.~H.}\ \bibnamefont {Tye}},\
  }\href {\doibase 10.1016/0550-3213(86)90362-7} {\bibfield  {journal}
  {\bibinfo  {journal} {Nucl. Phys.}\ }\textbf {\bibinfo {volume} {B269}},\
  \bibinfo {pages} {1} (\bibinfo {year} {1986})}\BibitemShut {NoStop}%
\bibitem [{\citenamefont {Mizera}(2017)}]{Mizera:2017cqs}%
  \BibitemOpen
  \bibfield  {author} {\bibinfo {author} {\bibfnamefont {S.}~\bibnamefont
  {Mizera}},\ }\href@noop {} {\  (\bibinfo {year} {2017})},\ \Eprint
  {http://arxiv.org/abs/1706.08527} {arXiv:1706.08527 [hep-th]} \BibitemShut
  {NoStop}%
\bibitem [{\citenamefont {Matsumoto}(1998)}]{matsumoto1998}%
  \BibitemOpen
  \bibfield  {author} {\bibinfo {author} {\bibfnamefont {K.}~\bibnamefont
  {Matsumoto}},\ }\href {https://projecteuclid.org:443/euclid.ojm/1200788347}
  {\bibfield  {journal} {\bibinfo  {journal} {Osaka J. Math.}\ }\textbf
  {\bibinfo {volume} {35}},\ \bibinfo {pages} {873} (\bibinfo {year}
  {1998})}\BibitemShut {NoStop}%
\end{thebibliography}

%

\end{document}